\documentclass[iop,apj]{emulateapj}
\usepackage{amsmath,amssymb}
\usepackage{graphicx}
\usepackage{natbib}
\usepackage{color}
\usepackage[colorlinks,citecolor=blue,linkcolor=red]{hyperref}
\usepackage{threeparttable}
\usepackage{rotating}

\def\s{{\rm\thinspace s}}
\def\km{{\rm\thinspace km}}
\def\Mpc{{\rm\thinspace Mpc}}

\def\kmps{\hbox{$\km\s^{-1}\,$}}
\def\kmpspMpc{\hbox{$\kmps\Mpc^{-1}$}}

\begin{document}

\title{Long Fading Mid-Infrared Emission in Transient Coronal Line Emitters:\\ 
        Dust Echo of Tidal Disruption Flare}


\author{Liming Dou\altaffilmark{1}, Ting-gui Wang\altaffilmark{1}, Ning Jiang\altaffilmark{1}, Chenwei Yang\altaffilmark{1}, 
Jianwei Lyu\altaffilmark{2} and Hongyan Zhou\altaffilmark{1,3} }
\altaffiltext{1}{CAS Key Laboratory for Researches in Galaxies and Cosmology,
University of Sciences and Technology of China, Hefei, Anhui 230026, China; doulm@mail.ustc.edu.cn, twang@ustc.edu.cn}
\altaffiltext{2}{Steward Observatory, University of Arizona, 933 North Cherry Avenue, Tucson, AZ 85721, USA}
\altaffiltext{3}{Polar Research Institute of China, 451 Jinqiao Road, Pudong, Shanghai 200136, China}

\begin{abstract}

The sporadic accretion following the tidal disruption of a star by a super-massive black hole 
(TDE) leads to a bright UV and soft X-ray flare in the galactic nucleus. The gas and dust surrounding 
the black hole responses to such a flare with an echo in emission lines and infrared emission. 
In this paper, we report the detection of long fading mid-IR emission lasting up to 14 years after 
the flare in four TDE candidates with transient coronal lines using the {\it WISE} public data release. 
We estimate that the reprocessed mid-IR luminosities are in the range between $4\times 10^{42}$ 
and $2\times 10^{43}$ erg~s$^{-1}$ and dust temperature in the range of 570-800K when {\it WISE} first 
detected these sources three to five years after the flare. Both luminosity and dust temperature decreases 
with time. We interpret the mid-IR emission as the infrared echo of the tidal disruption flare. 
We estimate the UV luminosity at the peak flare to be 1 to 30 times $10^{44}$ erg~s$^{-1}$ 
and for warm dust masses to be in the range of 0.05-1.3 M$_\sun$ within a few parsecs. Our results suggest 
that the mid-infrared echo is a general signature of TDE in the gas-rich environment.

\end{abstract}




\section{Introduction} \label{sec:intro}

If a star passes within the tidal disruption radius of a supermassive 
black hole, the tidal force exceeds the star's self-gravity. 
The star is disrupted, and about half of the stellar debris is subsequently 
accreted, leading to energetic flares with an integrated radiation of 
about 10$^{52}$ erg for a solar-type star \citep{Hills1975,Lidskii1979,
Rees1988, Rees1990}.This phenomenon is commonly known as a tidal disruption 
event (TDE). A few dozen TDEs or TDE candidates have been reported so far 
from mutli$-$wavelength surveys or serendipitous observations from X$-$rays 
to optical band (e.g., \citealp{Bade1996,Komossa1999,Komossa2008,
Gezari2009,Gezari2012,Levan2011,van2011,Arcavi2014,Holoien2014,Holoien2016,Alexander2016}; 
and see \citealp{Komossa2015} for a recent review). Theoretical works predict 
that TDE should produce a bright flare mainly in UV to soft X-ray bands, 
which decreases as a power law of $t^{-5/3}$, as fall-back debris, 
intersects and is accreted \citep{Evans1989,Phinney1989,Lodato2009,
Lodato2015}. The observed light curves can often be fitted with the above 
form fairly well. The optical light curve of PS1-10jh provides one 
of the best example for this \citep{Gezari2012,Gezari2015}. However, an 
exponential law may sometimes be described better in other cases 
such as ASASSN-14li \citep{Holoien2016}. 

In a gas-rich environment, the UV and soft X-ray continuum from TDEs 
will ionize and heat gas surrounding the black hole, giving rise to variable 
high ionization emission lines. Fading of strong high ionization coronal lines 
and brightening of [O III] emission on time scales of years were reported in a 
handful of star forming galaxies, which was interpreted as the light echo of TDEs 
\citep{Komossa2008,Wang2011,Wang2012,Yang2013}. The decline of the UV 
continuum was detected in archival GALEX observations of SDSS J0748+4712 
\citep{Wang2011}, and \citet{Palaversa2016} detected a dramatic fading in UV 
emission by a factor of four with follow-up {\it Swift} observations of SDSS J0952+2143. 
A detailed study of the variable UV emission, together with the optical light curve 
serendipitously observed by the LINEAR survey, suggested that the transient that 
powered the line emission was most likely a TDE \citep{Palaversa2016}.

Because gas is often mixed with dust, an echo in the infrared is also expected. 
\citet{Lu2016} made a detailed calculation of the light curves at different infrared 
bands for a simple spherical distribution of dust within 1 pc. They predicted that 
the dust emission, which peaked at 3-10 $\mu$m, has a typical luminosity between 10$^{42-43}$ 
erg s$^{-1}$ in the case of well studied TDE ASASSN-14li \citep[e.g.,][]{Miller2015,
Alexander2016,Holoien2016,van2016a}, if most UV light is reprocessed into infrared. 
We confirmed such a mid-IR echo in ASASSN-14li, which lags the detected optical
 flare by $\sim$ 36 days \citep{Jiang2016}, based on the {\it WISE} cryogenic and NEOWISE 
post-cryogenic survey (hereafter ALLWISE Release) and NEOWISE Reactivation Survey 
(hereafter NEOWISE-R) at 3.4 and 4.6 $\mu$m \citep[labeled as W1 and W2,][]{Wright2010,
Mainzer2011,Mainzer2014}; though, the luminosity in the infrared band is one to two orders of 
magnitude lower than the model prediction due to low dust content.

\begin{sidewaystable}
\caption{3.4 and 4.6\,$\mu$m magnitudes of the TDEs with ECLs\label{tbmag}}
\centering
\begin{threeparttable}
\begin{tabular}{c ccc|ccc|ccc|ccc}\\
\hline\hline
{Epoch}  & {MJD} & {$W1$} & {$W2$}  & {MJD} & {$W1$} & {$W2$}  & {MJD} & {$W1$} & {$W2$}  & {MJD} & {$W1$} & {$W2$}  \\
 {No.} &{}    &{mag} & {mag}   & {}    &{mag} & {mag} &  {}    &{mag} & {mag}   & {}    &{mag} & {mag} \\
     {} &    {(1)} &{(2)} &{(3)}    & {(1)} &{(2)} &{(3)} &{(1)} &{(2)} &{(3)} & {(1)} &{(2)} &{(3)}     \\
\hline
 &  \multicolumn{3}{c|}{SDSS J074820.66+471214.6}  &\multicolumn{3}{c|}{SDSS J095209.56+214313.3} & \multicolumn{3}{c|}{SDSS J134244.42+053056.1} & \multicolumn{3}{c}{SDSS J135001.49+291609.7} \\
\hline
$A_1$ & 55291.224 & 13.46$\pm$0.01 & 12.68$\pm$0.02 & 55324.442 & 13.61$\pm$0.01 & 12.57$\pm$0.02&  55210.739 & 13.35$\pm$0.02 & 12.18$\pm$0.01 & 55203.809 & 13.82$\pm$0.03 & 12.90$\pm$0.02  \\
$A_2$ & 55482.710 & 13.55$\pm$0.02 & 12.77$\pm$0.01 & 55515.335 & 13.76$\pm$0.02 & 12.76$\pm$0.02&  55385.827 & 13.39$\pm$0.01 & 12.30$\pm$0.02 & 55377.137 & 13.92$\pm$0.01 & 13.04$\pm$0.02 \\
$A_3$ & \nodata    &  \nodata       &   \nodata      & \nodata   &  \nodata       & \nodata       &  55573.424 & 13.53$\pm$0.01 & 12.52$\pm$0.02 & 55565.293 & 13.98$\pm$0.02 & 13.20$\pm$0.03 \\
$N_1$ & 56946.258 & 13.87$\pm$0.01 & 13.27$\pm$0.02 & 56788.936 & 14.04$\pm$0.02 & 13.39$\pm$0.03& 56671.059 & 13.74$\pm$0.01 & 13.01$\pm$0.03 & 56663.095 & 14.45$\pm$0.03 & 13.67$\pm$0.03 \\
$N_2$ & 57115.140 & 13.94$\pm$0.01 & 13.39$\pm$0.03 & 56979.690 & 14.07$\pm$0.03 & 13.44$\pm$0.04& 56850.610 & 13.79$\pm$0.02 & 13.09$\pm$0.03 & 56839.404 & 14.49$\pm$0.03 & 13.82$\pm$0.03 \\
$N_3$ & 57309.581 & 13.95$\pm$0.02 & 13.47$\pm$0.06 & 57148.249 & 14.00$\pm$0.02 & 13.37$\pm$0.04& 57037.070 & 13.81$\pm$0.01 & 13.15$\pm$0.02 & 57027.327 & 14.59$\pm$0.03 & 13.95$\pm$0.04 \\
$N_4$ & \nodata   &  \nodata       &   \nodata      & 57341.989 & 14.16$\pm$0.03 & 13.54$\pm$0.04& 57209.421 & 13.84$\pm$0.02 & 13.26$\pm$0.03 & 57199.612 & 14.61$\pm$0.03 & 13.96$\pm$0.04  \\
\hline
 & template  & 14.16$\pm$0.14 & 13.95$\pm$0.16 & template  & 14.27$\pm$0.22 & 14.35$\pm$0.20& template  & 13.47$\pm$0.30 & 13.23$\pm$0.35 & template  & 14.57$\pm$0.28 & 14.42$\pm$0.31 \\
\hline
      & $MJD_o$  & $z$            &                 & $MJD_o$   & $z$            &               & $MJD_o$   & $z$            &                 &  $MJD_o$ & $z$            &  \\
      & $(4)$    & $(5)$          &                 & $(4)$     & $(5)$          &               & $(4)$     & $(5)$          &                 & $(4)$    & $(5)$          &  \\
\hline
      & 53055  & 0.0615           &                 & 53734     &  0.0789        &               &  52373    &  0.0366        &                 & 53848    &  0.0777        & \\
\hline
\end{tabular}
\begin{tablenotes}
\item[] Columns (1) mean modified Julian date of the epoch; (2) mean magnitude in W1 band; (3) mean magnitude in W2 band; (4) modified Julian date of the SDSS spectroscopic observations; (5) redshift.
\end{tablenotes}
\end{threeparttable}
\end{sidewaystable}

Inspired by this discovery, we examine systematically the mid-infrared 
emission of the four TDE candidates with extreme coronal lines (ECLs) which  
strongly faded in \citet{Yang2013} using the newly released NEOWISE-R data (2016 March 
23), in combination with the ALLWISE data. Echo in the infrared light was 
indicated previously from the mid-infrared spectrum taken by {\it Spitzer} for 
J0952+2143 \citep{Komossa2009}. In this paper, we show that the mid-infrared emission has 
declined by about 0.5-1.1 mag between ALLWISE and NEOWISE-R surveys in the W1 and 
W2 bands for all four objects.

The paper is organized as follows. The data analysis and results are described 
in Section\, 2. We discussed the results and concluded in Section\, 3 and 4. Throughout 
this paper, we adopt a ${\Lambda}CDM$ cosmology with $\Omega_{M}$ = 0.3, 
$\Omega_{\Lambda}$ = 0.7, and a Hubble constant of $H_{0}$ = 70 $\kmpspMpc$.

\section{Data Analysis and Results} \label{sec:datares}

\subsection{The Mid-Infrared Light Curves }\label{sec:lc}

We extract the mid-infrared photometric data from the AllWISE Data Release 
and NEOWISE Reactivation Release for the four ECL emitters (ECLEs) within 
1\arcsec.0 offset to their optical positions. We downloaded and checked the 
{\it WISE} images for potential contamination. No other sources are found within 
10\arcsec. The signal-to-noise ratio (S/N) is high enough to give reliable 
photometric measurements with S/N $>$ 10 in W1 and S/N $>$ 5 in W2 for all 
exposures. There are $\sim$12 exposures in each epoch, provided by the 
unique-designed observational cadence of {\it WISE} for the detection of intraday 
variability. Using the same method as described in \citet{Jiang2012,Jiang2016}, 
we first examine the short-term variability, and find that all magnitudes 
within each epoch are consistent with being a constant. So we average these 
magnitudes to yield a mean value for each epoch when examining the long-term 
variability (Table\,\ref{tbmag}). The uncertainties of the mean is calculated 
following error propagation.
\begin{figure*}
\figurenum{1}
       \centering
        \begin{minipage}{0.85\textwidth}
        \centerline{\includegraphics[width=1.0\textwidth]{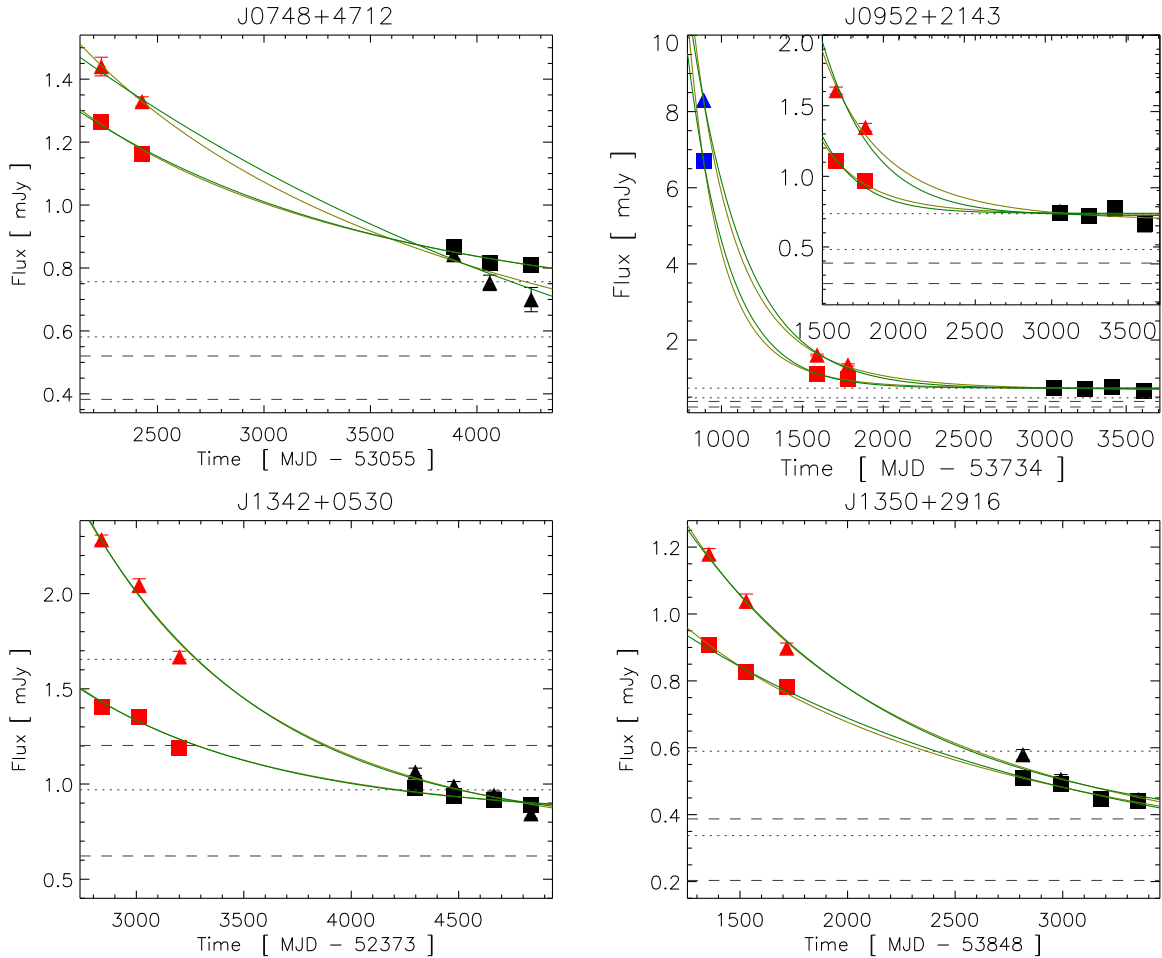}}
        \end{minipage}
        \caption{WISE light-curves in W1 and W2 bands and the empirical fits. The light-curve is fitted with an exponential law and a constant flux $f(t)=A e^{-Bt}+C$ (dark green lines); or a power-law and a constant flux $f(t)=A t^{-B}+C$ (dark yellow lines). The dark grey dot (dash) lines mark the 1-$\sigma$ upper and lower limit of galaxy magnitudes derived from the SED matching. The red data points are from ALLWISE catalog; the black data points are from the NEOWISE(-R) catalog, the blue data points of J0952+2143 are estimated from Spitzer. The time is referred to the date of SDSS spectroscopic observation. \label{fig:lcs}}
\end{figure*}

The dates of the optical spectroscopic observations from the SDSS database are 
listed in Table\,\ref{tbmag}. The start of the flare of J0952+2143 occurred 
approximately 580 days before the epoch of the SDSS spectroscopic observations 
\citep{Palaversa2016}. The continuum flares of J0748+4712 and J1350+2916 occurred 
within 120 and 700 days prior to the spectroscopic observations, respectively 
\citep{Wang2011,Wang2012}. Despite the fact that we do not have good 
constraints on the date for the continuum flare in J1342+0530, 
there are reasons to believe that the tidal disruption flare happened not 
too far from the SDSS spectroscopic observation. The fact that high ionization 
iron coronal lines disappeared in all subsequent observations suggested that 
they must be short lived. For J0748+4712, all coronal lines disappeared in the 
spectra taken at Xinglong and Lijiang station four to five years after the SDSS discovery 
\citep{Wang2012,Yang2013}. Some theoretical considerations would suggest an even 
shorter duration. In the cold interstellar medium, iron is mainly locked in the 
dust, so strong coronal lines are most likely produced in the region within 
the dust sublimation radius, which is a few light-months to a few light-years 
from the black hole, depending on the peak luminosity of the flare. 
In J0748+4712, the high coronal lines ([Fe XIV], [Ar XIV], [Fe XI] and [Fe X]) 
were detected within 4 months of the continuum peak, consisting with above 
estimate. The UV flux decreases exponentially with an e-folding time 
scale of a few months or in a power law t$^{-5/3}$, so the ionization 
parameter drops by a factor of 20-100 two years later, relative to 
120 days, depending on which law is used. This factor is large enough 
to make a transition to intermediate ionization gas (responsible for 
[Fe VII] lines). 
This gives an estimate of the interval time between the first {\it WISE} 
observation and the continuum flare. Therefore, we conclude that the 
continuum flares occurred five to nine years prior to the first IR epoch in each source.

The light curves in the W1 and W2 bands are presented in Figure\,\ref{fig:lcs}. 
In order to illustrate the IR decays, the time $t$ is referred to the epoch 
of the SDSS optical spectroscopic observation. Each curve consists of 
five to seven epochs in about 5.5 years with the first two or three from the 
ALLWISE release (red symbols; $A_i$, $i$ = 1, 2, 3 in table\, \ref{tbmag}) and 
the last three or four from the NEOWISE-R release (black symbols; $N_i$, $i$ = 1, 2, 
3, 4 in table\, \ref{tbmag}). There is a large gap ($\sim$2.8 years) between 
the two releases. The light curves looks remarkably similar for all objects. 
In both W1 and W2 bands, we observe a long term decline of the infrared emission. 
The sources were $\sim$ 0.5-1.1 mag brighter at the first epoch than at the 
last epoch in W1 and W2 bands; and the variability amplitude is larger in W2 than 
in W1. For J1342+0530, the ALLWISE catalog gives a variability flag of 7 and 9 for 
W1 and W2 bands, suggesting significant variability between different exposures 
during the ALLWISE survey. For other sources, the variability flags are all smaller 
than five, so variability among different exposures is not significant during ALLWISE 
survey. However, the mean flux of each epoch decreased significantly during the 
ALLWISE survey for all four sources according to $\chi^2$-test. This can be attributed 
to the increase of S/N over single exposures. We also detect significant decline 
within NEOWISE-R survey in three sources (J0748+4712, J1342+0530, and J1350+2916). 
J0952+2143 seems to display non-monotonous variability ($N_1$, $N_2$, and $N_3$ in 
table\, \ref{tbmag}) at 2$\sigma$ level, probably attributed to the complex 
dust distribution.

\begin{figure}
\figurenum{2}
\centering
 \begin{minipage}{0.42\textwidth}
       \centerline{\includegraphics[width=1.0\textwidth]{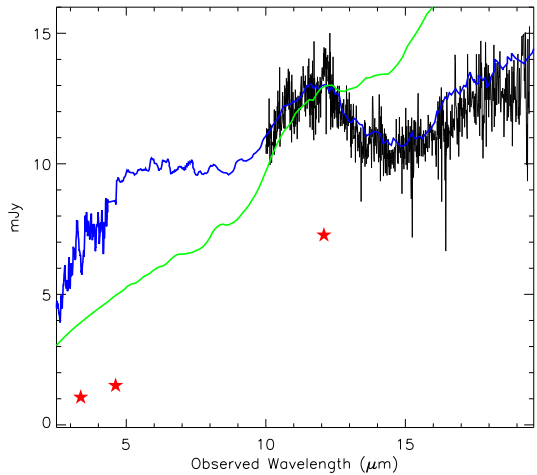}}
 \end{minipage}
 \caption{{\it Spitzer} SH IRS spectrum of J0952+2143. The blue line overplots the template of PG 1121+422 and the green line is an average template quasar SED of Netzer et al. (2007). The red pentacles are the W1, W2, and W3 flux from its $A_1$ epoch. \label{sed0952}}
\end{figure}

To extend the light curve, we estimate the W1 and W2 flux for J0952+2143 during 
the {\it Spitzer} observation. J0952+2143 was observed by {\it Spitzer}/IRS with the Short-High 
(SH) module on 2008 June 5 (MJD: 54624), which is $\sim$ 700 days before its 
first {\it WISE} epoch observation \citep[see][]{Komossa2009}. Since this object is 
unresolved in the SH aperture, we adopt its mid-IR spectrum with the optimal 
differential extraction from the Cornell Atlas of {\it Spitzer}/Infrared Spectrograph 
Sources \citep[CASSIS;][]{Lebouteiller2015}. The extraction of this spectrum was 
based on the BCD products processed by the final version of the {\it Spitzer} 
pipeline (S18.18). 
We derived a synthetic {\it WISE} W3 photometry from the {\it Spitzer} spectrum to 
be 12.0mJy, which is 1.8 times brighter than that at the A1 epoch.
We found that the mid-IR continuum of J0952+2143 is quite flat, and its mid-IR 
spectrum is almost the same as that of the hot-dust-deficient quasar PG 1121+422, 
which is composed of its 5-35 ${\mu}m$ {\it Spitzer} mid-IR spectra and 2.5-5 ${\mu}m$ AKARI 
near-IR spectra (Figure\, \ref{sed0952}; \citealt{Lyu2016}). 
Using the 2.5-35 ${\mu}m$ SED of the latter as a template, we derived the synthetic 
{\it WISE} W1 and W2 photometry of J0952+2143 to be 6.7 and 8.3mJy, which is $\sim$ 6.1 and 5.2 
times brighter than the one from the $A_1$ epoch. 
However, extreme care should be taken for such extrapolation, and we will discuss 
it in \S 2.3. It is interesting to note that mid-IR emission was fading out 
in all three bands, though with different amplitudes.

\subsection{Galaxy Contribution in the mid-IR} \label{sec:starlight}

The colors $0.78<W1-W2<1.17$ and $2.80<W2-W3<3.74$ on the first epoch are well in the 
locus of AGNs and LIRGs \citep{Wright2010}. The large-amplitude variability of the 
infrared flux on time scales of years rules out the starburst origin. Since these 
sources are radio quiet, the infrared emission cannot come from radio jets, rather 
they originate from dust heated by the accretion onto black holes. The long-term 
monotonous decline in all four objects and two mid-infrared band, suggesting that 
the primary accretion power decreases on the time scales of several years. Thus it 
is natural to associated the mid-infrared emission to the putative TDE flare, which 
also drives the fading coronal lines, similar to the infrared echo seen in 
ASASSAN-14li \citep{Jiang2016}. The dust within a few parsecs of the black hole, 
heated by the UV and soft-X-ray flare, re-emits in the mid-IR. 

To obtain the light curve for the TDE echo, we need to know the steady galaxy 
contribution in the W1 and W2 bands. First, we notice that the light curve becomes 
flatter with time, and seem to approach a stable flux at the end of NEOWISE-R. 
Second, $W1-W2$ decreases with time and moves into the range ($0.48-0.65$) of 
spiral galaxies on the last epoch (Wright et al. 2010). Therefore, the galaxy 
emission should be significant or even dominate W1 and W2 flux on the last epoch 
of NEOWISE-R. In the following, we will adopt two methods to estimate the background 
galaxy contribution.

In the first approach, we estimate the galaxy emission in the {\it WISE} W1 and W2 bands 
by using the SEDs constructed from SDSS and 2MASS photometry, which should be 
dominated by galaxy light except for J0952+2143, where we detected a significant 
contribution of the non-stellar component in the SDSS photometric flux within the SDSS 
fiber \citep{Wang2012}. We match these SEDs to the SDSS main galaxies with similar 
colors in optical and near infrared bands at redshifts ($0.02<z<0.1$), and $r$ 
magnitudes brighter than 17.5. For each TDE candidate, we found 10-100 matched 
galaxies. We extract {\it WISE} W1 and W2 magnitudes for these galaxies, and calculate 
the mean and standard deviation of $W1-K$ and $W2-K$ for the matched sample for 
each TDE candidate. We use the mean to estimate the W1 and W2 magnitude of the galaxy 
light and the standard deviation as the uncertainty. The estimated magnitudes and 
uncertainties are given in the Table\,\ref{tbmag}. This method works well for 
J0748+4712 with a scatter of 0.12 and 0.16 mag in W1 and W2 band, but for other 
sources, the typical uncertainty is still large (0.2-0.3 mag). 

\begin{table*}
\centering\scriptsize
\caption{Results of Mid-IR Light Curve Fit\label{lcfit}}
\begin{threeparttable}
\begin{tabular}{c cccc cccc c}\\
\hline
\hline
{Name}&$A_{W1}$& {$B_{W1}$} & {$C_{W1}$} & {$\chi^2_{W1}$} & {$A_{W2}$}& {$B_{W2}$} & {$C_{W2}$} & {$\chi^2_{W2}$} & {dof}  \\
{} & {} &{} &{mJy} &{} &{} &{} & {mJy}  &{} & {}   \\
{(1)} & {(2)} &{(3)} &{(4)} &{(5)} &{(6)} &{(7)} & {(8)} &{(9)} &{(10)} \\
\hline
 & \multicolumn{9}{c}{Model: $f(t)$ = A e$^{-B t}$ + $C$ }   \\
\hline
J0748+4712 & 3.04$\pm$1.45  & 0.25$\pm$0.10 & 0.66$\pm$0.11 &  5.7  & 3.08$\pm$0.10   & 0.12$\pm$0.01 & 0$\pm$*    & 4.1   & 2  \\
J0952+2143 &1882$\pm$142  & 1.43$\pm$0.02 & 0.74$\pm$0.01 &  38.7 & 600$\pm$39   & 1.09$\pm$0.02 & 0.72$\pm$0.01 &  25.2   & 3  \\
J1342+0530 &29.4$\pm$18.5  & 0.40$\pm$0.07 & 0.84$\pm$0.03 &  19.8 &103.7$\pm$41.4  & 0.43$\pm$0.04 & 0.76$\pm$0.05 & 19.2  & 4 \\
J1350+2916 & 2.15$\pm$0.46  & 0.20$\pm$0.07 & 0.20$\pm$0.14 &  2.9  & 6.24$\pm$1.52   & 0.36$\pm$0.05 & 0.34$\pm$0.05 & 13.4  & 4 \\
\hline
 & \multicolumn{8}{c}{Model: $f(t) = A t^{-B} + C$ }  \\
\hline
J0748+4712 &10.32$\pm$13.13  & 1.38$\pm$0.90 & 0.48$\pm$0.32  &  5.2  & 10.34$\pm$0.76 & 1.06$\pm$0.04 & 0$\pm$*    &  4.6   & 2    \\
J0952+2143 &105910$\pm$13530  & 7.03$\pm$0.09 & 0.73$\pm$0.01 &  28.5 & 12314$\pm$1376 & 5.3$\pm$0.08 & 0.68$\pm$0.02 &  9.1   & 3  \\
J1342+0530 &2903$\pm$4950  & 3.68$\pm$0.78 & 0.77$\pm$0.06 &   19.1 & 17848$\pm$19436   & 4.08$\pm$0.50 &  0.62$\pm$0.08    &  16.3  &4 \\
J1350+2916 & 5.85$\pm$0.36  & 1.08$\pm$0.03 &  0$\pm$*    &    3.6 &  20.9$\pm$11.0  & 1.74$\pm$0.36 &    0.13$\pm$0.12    & 10.6   & 4 \\
\hline\hline
\end{tabular}
\begin{tablenotes}
\item[]
Columns (1), SDSS name; (2)-(9), the best fit parameters from $f(t)=A e^{-Bt}+C$ or $f(t)=A t^{-B}+C$ and total $\chi^2$; (10), the degree of freedom (dof).
\end{tablenotes}
\end{threeparttable}
\end{table*}

In the second approach, we assume that the light curve of TDE echo can be described 
empirically with a simple analytic function and fit the light curves of W1 and W2 
with a combination of this function and a constant background. The light curves of 
the TDE in the UV and optical band can be described either as a power-law with an 
index of 5/3 \citep[$f(t) \propto t^{-5/3}$,][]{Gezari2009,Gezari2012} or as an 
exponential law empirically \citep{Holoien2014,Holoien2016}. The reprocessed infrared 
emission is the convolution of the UV light curve with the transfer function, determined 
by the dust distribution, in the case of thermal equilibrium.
The transfer function will cause two effects: a delay in the peak and smoothening the 
light curve. For a spherical shell of dust, the transfer function is a box function of 
width $2R/c$. At late times (t $\gg$ 2R/c), one can verify that the convolved light curve 
preserves the function form if the light curve can be described by either a power law or 
an exponential function. For very extended dust distribution, the flattening effect would 
still be significant even at late times. However, at a specific band, the light curve is 
further weighted by the Planck function. As temperature decreases to less than 800 K, 
the weights in W1 and W2 bands are decreased, so it has an opposite effect, i.e., 
steepening the light curve. As a result, it may be a good approximation to describe 
the infrared light curve using a power law or an exponential function.
We fit the light curves in W1 and W2 bands independently with a combination of either 
a power law or an exponential law and a constant flux: $f(t)=A t^{-B}+C$ 
or $f(t)=A e^{-Bt}+C$, where $C$ represents for the galaxy emission and $t$ is the time 
from TDE. Because the TDE time is not known for these sources except for J0952+2143, we 
use the best constrained time. Fortunately, the final result is not sensitive to the 
exact value within the constrained ranges.

The best-fitted model parameters are listed in Table\,\ref{lcfit}. Overall, the power-law 
fit is better than the exponential law fit, especially for J0952+2143 with $\Delta\chi^2=10.2$ 
and 16.1 for W1 and W2 with the same degree of freedom. However, in the power-law fit, the galaxy 
component $C$ is poorly determined except for J0952+2143 and J1342+0530, while it can be well 
constrained in exponential law except for W2 of J0748+4712. As shown in Table \,\ref{lcfit}, the 
best-fitted $C$s are consistent with those estimated from the SED matching, which have larger 
uncertainties except for J0748+4712. In the following analysis, we will use $C$ from the 
light curve modeling for objects other than J0748+4712. For the latter, we adopt the galaxy 
emission from the SED matching method.

\subsection{Dust Temperature and Mid-infrared Luminosity}\label{sec:lum}

After subtracting the galaxy emission, we estimate the reprocessed mid-IR flux due to the TDE flare in each epoch for each source at W1 and W2 band. Assuming a black-body spectrum, we calculate the dust temperatures and black-body luminosities (Table\,\ref{tbflux}). On the first epoch of {\it WISE} $A_1$, the temperatures are in the range of 560-870K and the logarithmic black-body luminosity in a narrow range from 42.7 to 43.1 (erg~s$^{-1}$). The uncertainties of the black-body luminosity and temperature are introduced mainly by the uncertainties in the galaxy contribution. For ALLWISE data, the uncertainty is 6-130 K in $T$ and 0.02-0.25 dex in $\log L_{bb}$. However, the temperature has relatively large uncertainties on the NEOWISE-R epochs.

For J0952+2143, we also estimate the dust temperature and black-body luminosity during the {\it Spitzer} observation using the same method. The estimated black-body luminosity ($\log L_{bb}$ = 44.05) during the {\it Spitzer} observation is $\sim$7.4 times higher than on the first {\it WISE} epoch. The uncertainty of the estimated galaxy emission of J0952+2143 is small. The dust temperature from {\it Spitzer} ($T = 980 K$) are also significantly higher than for other temperatures (607, 562, and 423 K) estimated from the first three {\it WISE} epochs. The last three temperatures are very uncertain due to the lack of precise galaxy contribution. The temperature decline follows a power law ($T(t)={\alpha} t^{-{\beta}}$). 
With {\it Spitzer} and the first three {\it WISE} temperatures, we obtain $\beta = 0.81\pm0.02$; while with only the first three {\it WISE} temperatures, 
we obtain $\beta = 0.68\pm0.20$, which is consistent with the previous one.

The extrapolation from the {\it Spitzer} spectrum to W1 and W2 has a large uncertainty. A better strategy would simultaneously analyze the W3 light curve. However, because of running out of cryogenic coolant, W3 is no longer available after 2010 September. Thus we have W3 photometry in only one epoch (A1). A detailed comparison between W3 flux at A1 epoch and the {\it Spitzer} spectrum is shown in the Figure 2. By converting {\it Spitzer} to W3 filter gives a flux that is 1.8 times as high as that of {\it WISE} flux at A1 epoch, while in the W1 and W2 bands, the interpolation by matching a hot-dust-deficient quasar template to the Spitzer spectrum gives a W1 and W2 flux five to six times higher than those of {\it WISE} at A1 epochs. Now we estimate the galaxy background at around 12 $\mu$m. Assuming that a single black-body describes the reprocessed infrared emission of TDE, using the W1 and W2, we obtain 2.2 mJy for the TDE component on A1, so the galaxy background is 4.5 mJy. Based on this estimate, the TDE component has varied by a factor of 3.4 at 12 $\mu$m from {\it Spitzer} observation to A1. After taking the background estimate in Table 2, the variability amplitude in W1 and W2 bands are factor of 15 and 8. 

\begin{figure}
\figurenum{3}
\centering
 \begin{minipage}{0.42\textwidth}
       \centerline{\includegraphics[width=1.0\textwidth]{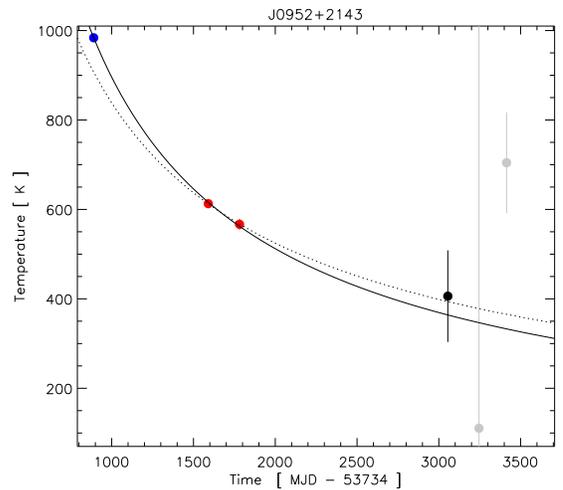}}
 \end{minipage}
 \caption{Variations of dust temperature in J0952+2143. The dust temperature decline follows a power law ($T(t)={\alpha} t^{-{\beta}}$).
Using the {\it Spitzer} and the first three {\it WISE} temperatures, we obtain $\beta = 0.81\pm0.02$ (solid line); while using only the first three {\it WISE} temperatures,
we obtain $\beta = 0.68\pm0.20$ (dotted line). \label{temp0952}}
\end{figure}

Now we make an independent estimate of variability in the W1 and W2 bands during the {\it Spitzer} observation based on the dust temperature evolution. 
In Figure\, \ref{temp0952}, we found that the dust temperature evolution can be described approximately by a power law with an index of 0.68 for the first three {\it WISE} data alone. If we extrapolate this temperature model to the {\it Spitzer} observation, we will get a dust temperature of 908$^{+175}_{-150}$ K, which is consistent with the matched quasar template within error. Note that at the epoch A1 the dust temperature is around 607 K. This means that
the flux decrease of the TDE component in W1 and W2 bands is a factor of 4.5 and 2.5 times larger than that in W3 band, which is roughly consistent with the estimate in the last paragraph.

\section{Discussion}\label{sec:dis}

We found a long-term and large-amplitude decline in the mid-IR flux
lasting up to 13 years after the first detection of coronal lines in four TDE candidates with ECLs.
 Using a power law and/or an exponential law, we fitted the light curves and estimated the flux 
 contributions from the host galaxies and warm dust heated by the TDE flares. 
 We found that the observed mid-IR luminosity is in range of $4\times 10^{42}$ to $10^{43}$ erg s$^{-1}$ when they were first caught by {\it WISE}. 

\subsection{The Peak UV Luminosity}\label{sec:bol}

Assuming an initial spherical distribution of dust around the black hole, as the continuum flare passed by, the dust grains within sublimation radius at the peak luminosity are heated to a high temperature and then evaporated. It is unlikely that dust will be formed again within $R_{sub}$ under the strong UV radiation of TDE when the radiation temperature falls below the sublimation temperature late on. This will leave a cavity of dust free region around the black hole. 

Grain temperature $T$ is determined by the balance between the heating by UV-optical photons and its thermal radiation \citep{Laor1993}: 
\begin{equation}
\int_0^\infty\frac{L_\nu}{4\pi r^2} e^{-\tau_\nu}Q_{abs}(\nu)d\nu =
\int_0^\infty 4\pi B(T_d,\nu)Q_{abs}(\nu)d\nu ,
\end{equation} 
where $L_{\nu}$ is the specific luminosity, $\tau$ is the optical depth, $Q_{abs}$ is the absorption efficiency, and $B(T_d,\nu)$ is the Planck function. For simplicity, we will assume that $\tau$ is small. In practice, one can express the integration on the right side in terms of 
\begin{equation}
Q(T_d)=\frac{\pi}{\sigma T_d^4}\int_0^{\infty}Q_{abs}(\nu)B(T_d,\nu)d\nu.
\end{equation}
Similarly, one can define the average absorption co-efficiency in UV:
\begin{equation}
Q_{UV}=\int_0^{\infty}Q_{abs}(\nu)L(\nu)d\nu/L.
\end{equation}
$Q_{UV}$ depends somewhat on the shape of the UV continuum. In the case of TDE, the UV continuum can be approximated with a black body of temperature in the range of a few $\times10^4$ K \citep{Komossa2015} with little evolution on time \citep[e.g.,][]{Holoien2016}. It is a good approximation to assume $Q_{UV}=1$ for grains with $a > 0.01 \mu$m. 

The dust sublimation radius can be written as
\begin{equation}
R_{sub}\simeq 0.06 Q(T_d)^{-0.5} \left(\frac{L_{UV, peak}}{10^{45}\,erg\,s^{-1}}
\right)^{0.5} \left(\frac{1800K}{T_{sub}}\right)^2 pc. 
\end{equation}

In the observed temperature range, $300<T_d<1000$K, silicate grains behave like a gray emitter, i.e., $Q(T_d)\simeq const$, while graphite grains show a more complex $Q(T_d)\propto T_d^\beta$ \citep[$\beta\simeq 1.5$,][]{Draine1984,Laor1993}. For gray grains, as the UV continuum decreases approximately with time as $t^{-5/3}$, the dust temperature at $R_{sub}$ decreases as $T(t)\simeq T_{sub} (t/t_0)^{-5/12}$, where $t_0=R_{sub}/c$ from the time of UV flare. For graphite grains, the temperature declines as $T(t)\simeq T_{sub} (t/t_0)^{-0.30}$. 

\begin{table*}
\centering
\caption{Dust temperature and luminosity\label{tbflux}}
\begin{threeparttable}
\begin{tabular}{c cc|cc|cc|cc}\\
\hline
\hline
{Epoch} & {T}   & {$\log L_{dust}$} & {T}   & {$\log L_{dust}$} & {T}   &{$\log L_{dust}$} & {T}   & {$log L_{dust}$}  \\
{No.}   & {K}   & {erg s$^{-1}$}   & {K}   & {erg s$^{-1}$}   & {K}   &{erg s$^{-1}$}   & {K}   & {erg s$^{-1}$}           \\
{}      & {(1)} & {(2)}             & {(1)} & {(2)}             & {(1)} &{(2)}             & {(1)} & {(2)}           \\
\hline
 &  \multicolumn{2}{c|}{J0748+4712}  &\multicolumn{2}{c|}{J0952+2143} & \multicolumn{2}{c|}{J1342+0530} & \multicolumn{2}{c}{J1350+2916} \\
\hline
\multicolumn{9}{c} { Single Temperature Blackbody Model}\\
\hline
$Spi$ & \nodata       &  \nodata         & 981$\pm$2 & 44.05$\pm$0.01   & \nodata       &  \nodata      & \nodata       &  \nodata  \\
$A_1$ & 796$\pm$122   & 42.90$\pm$0.22   & 607$\pm$9& 43.18$\pm$0.02   &  584$\pm$20  & 42.71$\pm$0.08 &  804$\pm$68  & 43.04$\pm$0.13 \\
$A_2$ & 756$\pm$131   & 42.85$\pm$0.25   & 562$\pm$12& 43.09$\pm$0.04   &  607$\pm$25  & 42.61$\pm$0.09 &  812$\pm$82  & 42.96$\pm$0.16 \\
$A_3$ &\nodata        &\nodata           & \nodata    & \nodata          &  599$\pm$35  & 42.47$\pm$0.13 &  892$\pm$116  & 42.87$\pm$0.18 \\
\hline
\multicolumn{9}{c} {Modified Blackbody Model ($\propto \nu^{1.5} B(\nu, T)$)}\\
\hline
$Spi$ & \nodata       &  \nodata        & 690$\pm$4 &43.97$\pm$0.01   & \nodata       &  \nodata      & \nodata       &  \nodata  \\
$A_1$ & 595$\pm$67   & 42.81$\pm$0.21   & 485$\pm$37 & 43.06$\pm$0.02   &  471$\pm$12   & 42.58$\pm$0.06 &  601$\pm$37  & 42.96$\pm$0.13 \\
$A_2$ & 574$\pm$71   & 42.76$\pm$0.22   & 456$\pm$42& 42.95$\pm$0.04   &  485$\pm$15   & 42.49$\pm$0.07 &  605$\pm$45  & 42.88$\pm$0.15 \\
$A_3$ &\nodata       &\nodata           & \nodata    & \nodata          &  480$\pm$21   & 42.35$\pm$0.09 &  647$\pm$61  & 42.79$\pm$0.18 \\
\hline\hline
\end{tabular}
\begin{tablenotes}
\item[] 
Columns. (1), dust temperature; (2), dust luminosity. 
\end{tablenotes}
\end{threeparttable}
\end{table*}

The dust exposed to the peak of continuum flare on average is at a distance of $R = ct$\footnote{The iso-delay surface is actually a paraboloid with the focus at the black hole and line of sight as the symmetric axis.}, and has a temperature of about $T\simeq T_{sub} (R/R_{sub})^{-0.5}=T_{sub}(t/t_0)^{-0.5}$ for gray dust and $T\simeq T_{sub}(t/t_0)^{-0.36}$ for graphite grains. We expect the mid-IR emission at time $t$ comes from the dust between the two boundaries. This analysis suggests that at any observing time, the average dust temperature in the inner region is only mildly higher ($(t/t_0)^{1/12}$ for gray dust or $(t/t_0)^{0.06}$ for graphite dust) than that in the outside region. So we can approximate the mid-IR emission with a single temperature black-body. Assuming that the amount {\bf of} dust increases outward ($\rho_d\propto r^{-\gamma}$ with a $\gamma<2$), the mid-IR emission would be dominated by the dust illuminated with the peak of the flare, which has a mean radius of $R\simeq ct$. In the optically thick case, the dust temperature from the outer part will be lower than the above value by about $\exp (-\tau/4)$ due to dust extinction, where $\tau$ is the absorption depth in UV. 

With the above approximation, we can estimate the bolometric luminosity at the flare peak from the black body temperature. We will adopt $Q(T_d)$ in \citet{Laor1993}: $Q(T_d)\simeq 0.03$ for silicate grains, and $Q(T_d)\simeq 0.03 (T_d/630)^{1.5}$ for graphite grains of typical size $a = 0.1 \mu$m.
To be consistent with the non-gray nature of graphite in the mid-IR absorption, we recalculate the temperature from $f_{W1}/f_{W2}$ using $\nu^{1.5} B(\nu, T)$ (Table\,\ref{tbflux}). The logarithmic bolometric luminosities are 44.6, 44.0, 44.3, and 44.4 (erg~s$^{-1}$), for J0748+4712, J0952+2143, J1342+0530, and J1350+2916, respectively, for graphite grains with $a = 0.1 \mu$m, and mildly larger for silicate grains. The peak luminosity is sensitive to the adopted grain size, and will be 0.5-1 dex higher for 1$\mu$m grains. Thus, it is only an order of magnitude estimate unless the grain size distribution can be independently inferred by modeling the infrared spectrum or/and the extinction curve. Broadly, these luminosities are well in the range of the observed value for non-jetted TDEs discovered so far \citep{Komossa2015}.

In a similar way, we estimate the logarithmic bolometric luminosity from {\it Spitzer} IRS data to be 44.5 (erg~s$^{-1}$) 
assuming graphite grains with $a = 0.1 \mu$m. The last value is a factor of two higher than the above estimate from {\it WISE} 
data. The large value in the latter may indicate a significant optical depth between the main emission region during the 
{\it Spitzer} observation to that during the {\it WISE} observation. Because the absorbed UV light will be re-emitted in infrared, the 
total energy in the infrared, emitted during the period from the {\it Spitzer} observation to the first {\it WISE} observation should 
be a significant fraction of total energy in UV. Since UV flux declines steeply, the main energy comes from the early 
few months. Assuming that this stage is a few months to a year, the integrated UV flux would be about 
$(3-10)\times 10^{51}$ erg. This number is indeed comparable to the total mid-IR emission during 
the period (see below). 
However, since W1 and W2 fluxes during the {\it Spitzer} observation are inferred indirectly by matching the IRS 
spectrum, which does not cover W1 and W2 bands, to the mid-infrared spectrum of a quasar, we warn that it is entirely 
possible that the hot-dust component may be different from the quasar.
Unfortunately, the dust temperatures during the NEOWISE-R have large uncertainties due to poorly determined background, 
otherwise, one can check the slope of temperature decline to constrain the dust optical depth with a more detailed model.

For J0952+2143, we calculate the total energy emitted in the infrared from its light curve. 
Integrating the black body luminosities in Table \ref{tbflux} from the {\it Spitzer} epoch to the 
last WISE epoch $A_2$ yields $3\times 10^{51}$ ergs. The steep decline light curve from {\it Spitzer} to {\it WISE}
 suggests that the bulk output in the mid-IR was emitted in the early time; lack of the mid-IR data at earlier
 epochs means that the integrated value in above is only a lower limit. This is not a small fraction of the total 
 energy ($\sim$0.05$M_\sun c^2 \simeq 10^{53}$ erg) release in the accretion process for tidal disruption of a 
 solar-type star, assuming that half of the debris material is accreted onto black holes and the radiation 
 efficiency ($\eta \equiv L/ \dot{m}c^2$) is 0.1. While the tidally disrupted star comes mainly from the lower main 
 sequence, with a typical mass $\sim$0.3 solar masses \citep{Stone2016}, the dust would be reprocessing 10\% 
 fraction of the TDE radiation. Additional, the total radiation energy may be much lower if strong outflow is 
 lunched in the super-Eddington accretion phase \citep[e.g.,][]{Strubbe2009,Alexander2016,Metzger2016}. This means 
 that both the covering factor and optical depth of dust is not very small.

\subsection{Dust mass}\label{sec:mass}

We estimate the dust mass responsible for the IR emission. Assuming spherical grains with a size 
distribution $n(a)\, \propto\, a^{-3.5}$ and $a_{min} = 0.005 {\mu}m$, $a_{max} = 0.3 {\mu}m$ \citep{Mathis1977}, 
similar to those in the MW or S/LMC \citep{Draine1984}, we can write 
\begin{equation}
L_{IR}\, {\simeq}\, 4{\pi}{\sigma}{T^4} \int^{a_{max}}_{a_{min}} n(a)\,a^2\, <Q(a,T)> da ,  
\end{equation}
\begin{equation}
M_{dust}\, {\simeq}\, {\frac{4}{3}}{\pi}{\rho}\, \int^{a_{max}}_{a_{min}} n(a)\,a^3\, da , 
\end{equation}
where $a$ and $\rho$ are the radius and the density of the grain, respectively; $\sigma$ is the Stefan$-$Boltzmann 
constant. Noting that 
\begin{equation}
<Q(T,a)> \simeq Q_0(T,a) (a/1{\mu}m) .
\end{equation} 
It is a good approximation $Q_0 \simeq 0.3 $ for silicate grains of sizes less than 1 $\mu$m, and $0.3 (T_d/630K)^{1.5}$ 
for graphite grains of sizes less than 0.1 $\mu$m for $250<T_d<1000$ K. $Q(T, a)$ increases with $a$ for graphite grains 
up to the peak at 0.3-0.6 $\mu$m, and then decreases. Then we can write 

\begin{eqnarray}
M_{dust} & \simeq & \frac{\rho L_{ir} (1{\mu}m)}{3\sigma T_d^4 Q_0(T_d)} \\
&\simeq &0.057 \frac{\rho}{2.5 g\,cm^{-3}} \frac{L_{ir,43}}{Q_0(T_d)}\left(\frac{600K}{T_d}\right)^4 M_\sun. 
\end{eqnarray}  

Assuming an average density of $\rho =2.5 g$ cm$^{-3}$ \citep[see section 7.3.5 in][]{Kruegel2003}, the dust mass 
is 0.14, 0.76, 0.29, and 0.18 $M_{\sun}$ for J0748+4712, J0952+2143, J1342+0530, and J1350+2916, 
respectively, using the values in Table \ref{tbflux} on $A_1$ epochs for the graphite grains with $a \lesssim$ 0.1 
$\mu$m. For the large graphite grains with $a \sim 1 \mu$m, the dust mass would be increased by 67\%. Changing to 
silicate grains, the dust mass would be reduced by $\sim$ 65\%. Assuming a dust to gas ratio of 0.01, the amount of gas 
is 14, 76, 29, and 18 $M_{\sun}$ for J0748+4712, J0952+2143, J1342+0530, and J1350+2916, respectively. 
This is consistent with the recent theoretical works on the circumnuclear medium in quiescent 
galaxies \citep{Generozov2015,Generozov2016}. Therefore, the dust and gas in the four TDEs with ECLs is about 1000-5000 
times richer than the one in ASASSN-14li \citep{Jiang2016}. 

In comparison, our Galaxy has a molecular torus or circumnuclear disk (CND) with a total gas mass of 
$5\times 10^4\,M_\sun$, an inner edge at 1.4 pc and the outer-edge extending to 4-7 pc around the super-massive 
black hole \citep{Morris1999,Etxaluze2011}. The amount of warm dust detected in our TDE candidates is several orders of 
magnitude lower than that in CND of our Galaxy. However, in the Galactic center, the CND consists of clumps of sizes of 
about 0.1-0.2 pc and density of $10^{5-6}$~cm$^{-3}$ \citep{Genzel1989,Martin2012,Lau2013}. The individual clump is thus 
optically thick to UV and optical radiation, so only the surface of these clumps in the inner edge of the CND would be 
heated if there were a TDE in the Galactic center. Thus, dust with an amount similar to that of CND in the Galactic 
center cannot be ruled out. \citet{Lau2013} inferred the edge of CND as a ring of a thickness 0.34 pc at a 
radius 1.4 pc, which correspond to a covering factor of 12\%.

\subsection{On the Relation with Coronal Line Emission}

It is interesting that our coronal line sample all show the signal of infrared echo more than ten years after the 
flare. The first WISE observations were all more than five years after the peak of the UV/optical burst, and lasted 
for at least another three years. In contrast to the short echos of ASASSN-14li, for which the mid-IR echo is 
detectable only in 36-220 days after its UV/optical flare \citep{Jiang2016}. Thus, the dust in TDEs with ECLs would 
be much farther and thicker than that of ASASSN-14li. This is expected because dust is often mixed with gas. As 
argued by \citet{Wang2012}, TDEs with ECLEs occur in gas-rich nuclei. It is puzzling that the gas so close to the 
black hole does not trigger noticeable nuclear activity in those objects. Spectroscopic observations of the source 
after the echo signature passed away would be interesting. Long-term monitor in mid-IR band as well as spectroscopic 
observation can probe the distribution and composition of the dust in the very center of quiescent galaxies.

We observed these sources on 2011 December 26, which in between the {\it WISE} and NEOWISE-R phases \citep{Yang2013}, with the 
MMT telescope. We detected reprocessed emission lines in all four objects as indicated by either enhanced or fading 
[Fe VII] lines, or enhanced [OIII] lines. However, the infrared flux, estimated from the best-fitted light curve, 
does not correlate with the line flux of either [O III] or [Fe VII]. These differences may be due to the 
different geometry of the gas and dust, or light travel time effects. Because of the poor sampling and limited 
dynamic range of either parameter, we cannot give a good constraint yet. Among the four objects, J0748+4712 is the 
only one that does not show [Fe VII] emission lines in its MMT spectrum. However, neither its mid-IR infrared emission 
is the weakest one nor the dust temperature is the lowest in the sample. This indicates that gas ionization does not 
have any correlation with the equilibrium radiation temperature. 

\section{Conclusion }\label{sec:concl}

We detect a long-term and large-amplitude decline in the mid-IR light curves of four ECLs which are TDE candidates 
up to more than 10 years after the TDE using public {\it WISE} data archive. After subtracting the galaxy contribution, the 
light curves can be modeled with a power law and/or an exponential law. The fading mid-IR emission is naturally 
interpreted as the reprocessed UV TDE flare by dust in the inner several parsecs of the galactic nuclei. 
Our main results are as follows.

\begin{enumerate}
\item The reprocessed mid-IR luminosity lies in a relatively narrow range from $4\times 10^{42}$ to 
$2\times 10^{43}$ erg s$^{-1}$, and dust temperature of 460-600 K for graphite grain or 570-800 K for silicate grains 
during the first {\it WISE} observation, three to five years after the detection of coronal lines in the SDSS spectra. The {\it Spitzer} 
spectrum of J0952+2143 taken approximately two years before the {\it WISE} observation suggests a factor of five higher mid-IR luminosity, 
indicating the peak mid-IR luminosities may be much higher.

\item For J0952+2143, the integrated energy in the mid-IR is $3\times 10^{51}$ erg during the period between the 
{\it Spitzer} epoch and $A_2$ epoch. Due to the lack of the mid-IR at earlier epochs, we cannot give a good constraint on the 
total reprocessed energy for this object and other sources. 

\item The peak UV luminosity of the flare is estimated to be 1 to 30 times of $10^{44}$ erg~s$^{-1}$, well in the 
range of known TDEs.

\item The mass of warm dust is estimated to be 0.05-1.3 M$_\sun$ depending on the grain composition. The warm dust could 
be the skin of optically thick dusty clumps exposed to the UV flare or the inner edge of the dusty torus, as seen in the 
Galactic center.
\end{enumerate}

The {\it WISE} and NEOWISE-R data affords us an excellent opportunity to study the dust reverberation effect of TDEs 
at the infrared band (see also our work \citealp{Jiang2016}, and the recent work of \citealp{van2016b}). 
Using the full released 
{\it WISE} and NEOWISE-R data, we are examining such dust reverberation effects in all known TDEs systematically. It should be 
pointed out that there is still a large uncertainty in the background galaxy contribution, which affects the exact 
light curve and the estimate of dust temperature, particularly on the late epochs. Future mid-IR photometry should solve 
this problem because the TDE echo fades away. The estimated peak UV luminosity and dust mass depends on the grain 
properties. The mid-infrared spectrum can be used to constrain these properties, and thus to yield better constraints on the 
UV luminosity and warm dust mass. It is worth mentioning that if one can derive the dust extinction curve from UV and 
optical spectrum of a TDE, it can be used to constrain grain properties as well. Future mid-IR monitoring of newly discovered 
TDEs in infrared as well as in optical will yield very useful information about the SED of TDEs in the unseen UV, and 
explore the dust and gas in the inner parsecs of the quiescent galaxy.

We are grateful to the anonymous referee for comments and suggestions that have improved the quality of this paper.
We thank Wenbin Lu and Nicholas Stone for helpful discussions and comments.
This research is supported by the National Basic Research Program of China (grant No.\, 2015CB857005), the Strategic Priority Research Program ``The Emergence of Cosmological Structures" of the Chinese Academy of Sciences (XDB09000000), NSFC (NSFC-11233002, NSFC-11421303) and Joint Research Fund in Astronomy (U1431229) under cooperative agreement between the NSFC and the CAS, the Fundamental Research Funds for the Central Universities. The research of Yang C. is supported in part by the CAS ``Light of West China" Program (2015-XBQN-B-5).

This research makes use of data products from the {\it WISE}, which is a joint project of the University of California, Los Angeles and the Jet Propulsion Laboratory/California Institute of Technology, funded by the National Aeronautics and Space Administration. This research also makes use of data products from NEOWISE, which is a project of the Jet Propulsion Laboratory/California Institute of Technology, funded by the Planetary Science Division of the National Aeronautics and Space Administration.


\end{document}